\documentclass[preprint,showpacs,preprintnumbers,amsmath,amssymb,natbib]{revtex4}

\usepackage{graphicx,epsfig,dcolumn,bm}

\newcommand{\h} {\cfrac{\dot a}{a}}                                       
\newcommand{\hd}{\cfrac{\ddot a}{a}}                                      
\newcommand{\hh}{\left(\cfrac{\dot a}{a}\right)^2}                        
\newcommand{\st}{(x^i,p_j,\eta)}                                         
\newcommand{\ns}{(x^i,n_j,\eta)}                                         
\newcommand{\D}[2]{\cfrac{\mathrm{d} #1}{\mathrm{d} #2}}                 
\renewcommand{\d}[2]{\cfrac{\partial #1}{\partial #2}}                   
\newcommand{\pr}[1]{\left(#1\right)}                                     
\newcommand{\pc}[1]{\left[#1\right]}                                     
\newcommand{\ph}[1]{\left\{#1\right\}}                                   

\begin{document}

\title{Reproducing neutrino effects on the matter power spectrum through a degenerate Fermi gas approach}
\author{E. L. D. Perico}
\email{elduarte@ifi.unicamp.br}
\affiliation{Instituto de F\'{\i}sica Gleb Wataghin, Universidade Estadual de Campinas, PO Box 6165, 13083-970,
Campinas, SP, Brasil}
\author{A. E. Bernardini}
\email{alexeb@ufscar.br, alexeb@ifi.unicamp.br}
\affiliation{Departamento de F\'{\i}sica, Universidade Federal de S\~ao Carlos, PO Box 676, 13565-905, S\~ao Carlos, SP, Brasil}

\date{\today}

\begin{abstract}
Modifications on the predictions about the matter power spectrum based on the hypothesis of a tiny contribution from a degenerate Fermi gas (DFG) test-fluid to some dominant cosmological scenario are investigated.
Reporting about the systematic way of accounting for all the cosmological perturbations, through the Boltzmann equation we obtain the analytical results for density fluctuation, $\delta$, and fluid velocity divergence, $\theta$, of the DFG.
Small contributions to the matter power spectrum are analytically obtained for the radiation-dominated background, through an ultra-relativistic approximation, and for the matter-dominated and $\Lambda$-dominated eras, through a non-relativistic approximation.
The results can be numerically reproduced and compared with those of considering non-relativistic and ultra-relativistic neutrinos into the computation of the matter power spectrum.
Lessons concerning the formation of large scale structures of a DFG are depicted, and consequent deviations from
standard $\Lambda$CDM predictions for the matter power spectrum (with and without neutrinos) are quantified.
\end{abstract}

\date{\today}
\maketitle


\section{Introduction}

The cosmic accelerated expansion related to the conception of dark energy, the nature of dark matter components, the role of cosmological background neutrinos, and finally, the comprehension of all their intrinsic relations belong to one of the most challenging current problems in theoretical physics.
Such an enlarged overview of the dark sector does necessarily involve a fruitful interplay between general relativity, astrophysics and particle physics \cite{Teo2,Teo3,Teo4,Teo1}, which is fundamental to the linear perturbation theory that describes the cosmic inventory.

Besides presenting a consistent theoretical foundation, the theory for cosmological perturbations has been highly predictive.
In particular, it has explained the precise measurements of temperature and polarization anisotropies of the cosmic microwave background (CMB), most notably those from the Wilkinson Microwave Anisotropy Probe \cite{Bennet}.
Even being a matter of the same scope, the fine-tuning between theoretical predictions and observable data for large scale structures deserves, however, a sharper analysis.
In fact, it falls on more complex procedures for which even the nonlinearity effects are sometimes relevant \cite{Astropj}.
The matter power spectrum describes the density contrast of the Universe, which parameterizes the inhomogeneities in the matter distribution of the Universe.
The density contrast corresponds to the difference between the local energy density and the averaged energy density of some component of the cosmic inventory.
On large scales, the gravity competes with the cosmic expansion, and the structures grow according to the linear theory.
On small scales, the gravitational collapse is non-linear, and the accurate behaviour of the density contrast can only be computed by using N-body simulations.

Besides its intrinsic phenomenological character related to the large scale structures, the correct interpretation of the matter power spectrum is concerned with one of the experimental techniques for determining the mass of neutrinos through cosmological measurements, namely the CMB results.
It is done through the inference of a transfer function in the matter power spectrum at small scales \cite{Dodelson,Dolgov02}.
The contribution due to massive neutrinos to the closure fraction of cold dark matter at present can be depicted from the modifications on the matter power spectrum, even for neutrinos behaving like hot dark matter at higher redshifts.
It follows that the amount of cold dark matter at earlier epochs should be substantially reduced, suppressing the formation of large scale structures, when it is compared to a situation without massive neutrinos \cite{Boy10,Ma94,Pas06}.
The effects, however, are attenuated in case of hot dark matter or massless neutrinos.
A comparison of the observationally-inferred matter power spectrum with the power spectrum expected without the effects of massive neutrinos allows one to compute the effective contribution of neutrinos to the cosmic inventory.

The point at our manuscript is that the hypothesis of a tiny fraction of the cosmic inventory evolving cosmologically as a degenerate Fermi gas (DFG) test-fluid can mimic the massive neutrino cosmological behaviour at several dominant cosmological backgrounds.
The approach for treating the cosmological perturbation for ultra-relativistic neutrinos as hot dark matter, in general, differs from that for treating non-relativistic neutrinos as an additional cold dark matter component.
Treating massive neutrinos as a DFG not only gives novel ingredients to the recursive idea of a self-gravitating Fermi gas model but also allows for quantifying the smooth transition between ultra-relativistic (UR) and non-relativistic (NR) thermodynamic regimes for neutrinos in the cosmic background.
In a previous issue \cite{Ber10}, an analytical procedure based on Bianchi identities was adopted for obtaining the evolution of perturbations for a class of fluids which evolve from relativistic to non-relativistic regimes, a supposition that is usually adopted for cosmological neutrinos.
Our proposal becomes a convenient tool for describing the role of such particles in the cosmic inventory so that some peculiar modifications on the matter power spectrum can be debugged.

Assuming that the neutrinos temperature at present should be of about $T_{0\nu}\sim 10^{-4}K$ (depending on some phenomenological correspondence with the neutrino mass), and comparing this with a tiny fraction of a DFG test-fluid \cite{Ber10} in equilibrium with the radiation background in the beginning of the radiation-dominated era \cite{Bil97}, one should expect that the DFG's contribution to the matter power spectrum could be of the same order of magnitude of the neutrino's contribution.

Besides theoretical speculations about the existence of some kind of DFG produced in the early universe \cite{Zel81,Chandra1939,Kremer2002}, and the idea of a self-gravitating Fermi gas model introduced to explain the puzzling nature of white dwarf stars \cite{Fowler1926}, it may be expected that at some stage of the evolution of the Universe, primordial density fluctuations have become gravitationally unstable forming dense lumps of dark matter.
Such possibilities of forming large scale structures thus stimulate our meticulous investigation of the behaviour of a DFG as a test-fluid in the cosmological background of radiation, matter and cosmological constant ($\Lambda$).
Despite not necessarily resulting in matter lumps, we shall notice that it somehow contributes to the matter power spectrum related to large scale structures.

To quantify some analytical aspects related to the here proposed DFG approach, which results in fiducially modified and numerically confirmed predictions for the matter power spectrum of the cosmic inventory, our manuscript was organized as follows.
In section II we report about the formalism of linear perturbations focused on the study of a perfect fluid through the Einstein equation.
In section III we obtain some analytical expressions for the density fluctuation, $\delta$, and for the fluid velocity divergence, $\theta$, in case of a DFG at radiation-dominated, matter-dominated, and $\Lambda$-dominated background scenarios, and for super horizon solutions in case of a radiation-to-matter transitory regime.
In particular, we consider the conservation equation for the stress-energy tensor with a vanishing anisotropic stress tensor ($\sigma$).
Reporting about the systematic way of accounting for all the cosmological perturbations, i. e. through the Boltzmann equation, in section IV we quantify the behaviour of the inhomogeneities in the Fermi gas at the radiation-dominated era, through an ultra-relativistic analytical approach, and at the matter-dominated and $\Lambda$-dominated eras, through a non-relativistic analytical approach.
All the analytical results are consistently verified through the numerical calculations described in section V, where the DFG contribution to the matter power spectrum is compared with previously obtained numerical solutions for massive neutrinos corresponding to a tiny fraction of the dark matter components.
We draw our conclusions in section VI.

\section{Cosmological background and linear perturbations}

The cosmological evolution of a homogeneous Friedmann-Robertson-Walker (FRW) flat universe with averaged energy density, $\bar{\rho}(\eta)$, and averaged pressure, $\bar{\mathcal{P}}(\eta)$, is described in terms of the scale factor, $a(\eta)$, through the following components of the Einstein equation,
\begin{eqnarray}
\label{friedmann01}
\left( {da/d\eta \over a} \right)^{2} &=& {8\pi\over3}G a^{2} \bar{\rho} \,,\\
{d\over d\eta} \left( {da/d\eta \over a} \right)&=& -{4\pi\over 3}G a^{2} (\bar{\rho} + 3 \bar{\mathcal{P}}) \,,
\label{friedmann02}
\end{eqnarray}
where $\eta$ is the conformal time, $G$ is the Newtonian constant, and we have set the light velocity equals to unity, i. e. $c = 1$.

From the properties of a perfect fluid, the above elementary textbook equations can be depicted from the conservation properties of a stress-energy tensor $T^\mu_\nu$ that, at a comoving frame, is given by \cite{Ma94}
\begin{equation}\label{T_ideal}
T^{\mu\nu}=\mathcal{P} g^{\mu\nu}+(\rho+\mathcal{P})U^{\mu}U^{\nu}\,,
\end{equation}
where $U$ is the 4-velocity of the fluid particles, $g_{\mu\nu}$ is the metric tensor, and energy density and pressure are decomposed into averaged and perturbative values as $\rho = \bar{\rho} + \delta \rho$ and $\mathcal{P} = \bar{\mathcal{P}} + \delta \mathcal{P}$.
Once one assumes the validity of the Einstein equation, the corresponding conservation equation for a non-interacting single-particle fluid should be given in terms of the covariant derivative, $T^{\mu\nu}_{;\mu} = 0$.
One shall notice that the collective behaviour prescribed by our DFG approach guarantees that the stress-tensor is isotropic and diagonal, and that the energy-momentum tensor can thus be analytically described in terms of the abovementioned space-time dependence.

Throughout our study, we have chosen to work out the perturbation equations for the cosmological scenario in the longitudinal gauge since it has set simpler and clearer connections to preliminary studies on cosmological perturbations \cite{Boy10,Ma94,Pas06,Dodelson,Bernardini2011}.
By considering that the non-diagonal metric perturbations vanish \cite{Pas06}, two scalar degrees of freedom are eliminated from Eq.~(\ref{T_ideal}), and the remaining ones are defined in terms of two scalar potentials, $\phi$ and $\psi$, as
\begin{equation}
\label{longitudinal}
g_{\mu0}=-a^2(1+2\psi)\delta_{\mu0}\,, \qquad g_{ij}=a^2(1-2\phi)\delta_{ij}\,.
\end{equation}

The Bianchi identities \cite{Ma94} then provide us with the constraints on the stress-energy tensor in the Fourier $k$-space, which can be summarized by the continuity equation,
\begin{equation}
\label{dot_delta}
 \dot\delta=-(1+\omega)(\theta-3\dot\phi)-3\h (1+\delta)(1+\omega)-\frac{\dot{\bar\rho}}{\bar\rho}(1+\delta)\,,
\end{equation}
and by the Euler equation,
\begin{equation}
\label{dot_theta}
 \dot\theta=\frac{\,k^2\,\delta}{1+\omega}\frac{\delta \mathcal{P}}{\delta\rho} - \sigma \,k^2\, -4\h\theta + \psi \,k^2\, -\pr{\frac{\dot{\bar\rho}+\mathcal{\dot{\bar P}}}{\bar\rho}}\frac{\theta}{1+\omega}\,,
\end{equation}
with
\begin{equation}
\label{conser_cero}
a \d{\bar\rho}{a}+ 3 \bar\rho (1 + \omega) = 0 \,,
\end{equation}
where $\omega = \bar{\mathcal{P}}/\bar{\rho}$, and $\theta$, $\delta$ and $\sigma$ are, respectively, the density fluctuation, the fluid velocity divergence, and the shear stress defined by
\begin{equation}
\delta\equiv\delta\rho/\bar\rho, ~~~~  (\bar\rho+\mathcal{\bar P})\theta\equiv ik^j\delta T^0_j, ~~~~\mbox{and} ~~~~
(\bar\rho+\mathcal{\bar P})\sigma\equiv -\left(\hat k^i\hat k_j-\frac{1}{3}\delta^i_j\right)\Sigma^j_i,~
\end{equation}
with $\Sigma^j_i\equiv T^j_i - (\delta^j_i/3) T^k_k$.

The solutions of Eq.~\eqref{friedmann01} for radiation- and matter-dominated universes, and for the cosmological constant, with the corresponding equation of state respectively represented by $\mathcal{\bar P}_{\gamma} = \bar\rho_{\gamma}/3$, $\mathcal{\bar P}_m = 0$ and $\mathcal{\bar P}_\Lambda = -\bar\rho_\Lambda$, are given by
\begin{equation}
\label{rho(a)}
 \bar\rho_{\gamma} = \rho_0 \frac{\Omega_{\gamma}}{a^4}\,,
\qquad
 \bar\rho_m = \rho_0 \frac{\Omega_m}{a^3}\,,
\qquad
\mbox{and} \quad
 \rho_\Lambda=\rho_0\, \Omega_\Lambda\,,
\end{equation}
where the scale factor during an approach for radiation-to-matter transition era is given by,
\begin{equation}
 a(\eta)\approx \frac{2\pi G\rho_0}{3}\Omega_m\eta^2 + \sqrt{\cfrac{8\pi G\rho_0}{3}\Omega_{\gamma}}\eta \,,
\label{a(eta)}
\end{equation}
and during the $\Lambda$-dominated era by
\begin{equation}
a(\eta)\approx\pc{3\pr{\frac{\Omega_\Lambda^{1/3}}{\Omega_m^{1/3}}-\frac{4\sqrt{\Omega_{\gamma}\Omega_\Lambda}}{\Omega_m}}-\sqrt{\frac{8\pi G\rho_0\Omega_\Lambda}{3}}\eta}^{-1}\,.
\label{a(eta)2}
\end{equation}
where $\rho_0$ is the Universe's density at present, and $\Omega_s = \bar\rho_s/\rho_0$, with $s = \gamma,\, m,\, \Lambda$.

Finally, by decomposing the Einstein equation into time-time, longitudinal time-space, trace space-space, and longitudinal traceless space-space components, one obtains four equations to the linear perturbations in the $k$-space, which through Eq.~\eqref{longitudinal}, in terms of $\phi$ and $\psi$, are given by
\begin{subequations}
\label{einstein}
\begin{equation}
\label{E_tt}
 8\pi a^2G T^0_0 = -3\hh + 6\h\dot\phi+6\hh\psi + 2\,k^2\, \phi\,,
\end{equation}
\begin{equation}
\label{E_l_st}
 4\pi G a^2 (\bar\rho+\mathcal{\bar P})\theta=\,k^2\,\left(\dot\phi+\frac{\dot a}{a}\psi\right)\,,
\end{equation}
\begin{equation}
\label{E_tr_ss}
 \frac{4}{3}\pi G a^2T^i_i=-\hd+\frac{1}{2}\hh+\frac{\,k^2\,}{3}(\phi-\psi)+\ddot \phi+ \h(\dot\psi+2\dot\phi) + 2\psi \hd-\psi\hh\,,
\end{equation}
\begin{equation}
\label{E_l-tr}
 12\pi G a^2 (\bar\rho+\mathcal{\bar P})\sigma=\,k^2\,(\phi-\psi)\,.
\end{equation}
\end{subequations}

In the limit of ``single-particle'' fluid eras separately described by each element of Eq.~(\ref{rho(a)}), and in the absence of the shear stress, i. e. with $\sigma \approx 0$,  one can set $\psi \approx \phi$ in order to derive some analytical expressions for the evolution of the metric perturbation through scale factor dependence on $\eta$ given by Eqs.~(\ref{a(eta)}) and (\ref{a(eta)2}).
The full continuity, Euler and Einstein equations thus result in
\begin{subequations}\label{sol_phi}
\begin{equation}
\label{sol_rad}
\phi_{\gamma}(\eta) =  C_1 \frac{\cos\omega + \omega\sin\omega}{\omega^3} + C_2 \frac{\sin\omega - \omega\cos\omega}{\omega^3}\,,
\end{equation}
\begin{equation}
\label{sol_mat}
 \phi_m(\eta) = - \frac{D_1}{5\omega^5} + D_2\,,
\end{equation}
\begin{equation}
\label{sol_const}
\phi_{\Lambda}(\eta) = F_1(b-\eta)^3 +F_2(b-\eta)\,,
\end{equation}
\end{subequations}
with $\omega\equiv k\eta/\sqrt 3$.
The coefficients with sub-index $1$ in the above equations are related to the so-called isentropic decaying modes,
in the same fashion that the coefficients with sub-index $2$ are related to the isentropic growing modes.
As expected, decaying modes are suppressed during the cosmological evolution.
In addition, the solution for $\phi$ at the super-horizon limit during the radiation-to-matter transitory period is described by
\begin{equation}
\label{phi_Super}
\phi(y) = S_1\frac{\sqrt{1 + y}}{y^3} + S_2\frac{16 + 8 y - 2 y^2 - 9 y^3 }{y^3}\,,
\end{equation}
where $y\equiv a/a_{eq}=\bar\rho_m/\bar\rho_{\gamma}$, $a_{eq}$ is the scale factor at the time of equality, i. e. when $\bar\rho_{\gamma} = \bar\rho_m$, and $a \equiv a(\eta)$ is given by Eq.~(\ref{a(eta)}).

\section{DFG test-fluid solutions}

To obtain analytical solutions for a DFG as a test-fluid in each one of the cosmological background scenarios briefly discussed in the previous section, one has to follow a sequence of consistent simplifications.
Let us consider that the phase space distribution of the particles gives the number of particles in a differential volume $dx^{1} dx^{2} dx^{3} dp_{1} dp_{2} dp_{3}$ in phase space,
\begin{equation}
f(x^{i},\, p_{j},\, \eta)\,dx^{1} dx^{2} dx^{3} dp_{1} dp_{2} dp_{3}/(2\pi)^3 = dN,
\end{equation}
with $p_{j} \equiv p_{j}(p,\,n_{j})$, and where $f$ is a Lorentz scalar that is invariant under canonical transformations.
The general expression for the stress-energy tensor, in case of a single-particle fluid, written in terms of the distribution function $f$ and of the physical quadrimomentum components, $P^{\mu}$, is then given by
\begin{equation}
\label{T_General}
T_{\mu\nu}=\int \frac{dP_1 dP_2 dP_3}{(2\pi)^3(-g)^{1/2}} \frac{P_\mu P_\nu}{P^0} f\st\,,
\end{equation}
The phase-space distribution function evolves according to the Boltzmann equation as
\begin{equation}
\label{B}
\frac{Df}{d\eta} = \d{f}{\eta}+\D{x^i}{\eta}\d{f}{x^i}+\D{q}{\eta}\d{f}{q}+\D{n_i}{\eta}\d{f}{n_i}=\left(\d{f}{\eta}\right)_C\,,
\end{equation}
where, for non-interacting fluids, the right-hand (collision) term vanishes.
The fluid temperature is, however, not uniform.
Thus one conveniently sets $T = \bar T + \delta T$ with $\delta T = \delta T\ns$, $a \, p^i\equiv q^i \equiv q\,n^i$, where $q$ is the comoving momentum, and the time-dependence is relegated solely to $\bar T$.
Expanding the distribution function, $f(x^i, p_{j}(p, n_{j}),\eta)$ up to first-order in the temperature perturbation, $\delta T$,  with $f|_{\bar{T}} = \bar{f}(p, \bar{T}(\eta))$, one has
\begin{equation}
\label{Psi_dfs_pre}
f(x^i,p_j,\eta) \approx\left.f\right|_{\bar T}+\left.\d{f}{T}\right|_{\bar T}\delta T\ns \equiv \bar f(p,\eta)\pc{1+\Psi(x^i,p_j,\eta)}\,,
\end{equation}
and the Boltzmann equation in the Fourier $k$-space for the longitudinal gauge becomes
\begin{equation}
\label{PSI_dot_general}
\d{\bar f}{\eta}(1+\Psi)+\dot\Psi\bar f + i\frac{q}{\epsilon}(\vec{k} \cdot \hat{n})\Psi\bar f + \left(q\dot\phi-i\epsilon\,(\vec{k} \cdot \hat{n}) \psi\right)\d{\bar f}{q}=\left(\d{f}{\eta}\right)_C\,,
\end{equation}
where energy and momentum are respectively rewritten in terms of the (pseudo)comoving energy, $\epsilon = a\, E = a\, (p^{2}+m^{2})^{1/2} =  (q^{2} + a^{2}m^{2})^{1/2}$, and of the comoving momentum, $q = p\,a$.
By reducing the notation for the averaged temperature, $\bar{T}$, to $ \sim T$, the distribution function for a DFG can thus be given by
\begin{equation}
\bar f(p,\eta)= g_d \pr{\mathrm{exp}\pc{\frac{E-\mu}{T}}+1}^{-1}\approx
\left\{\begin{array}{ccc}
g_d & \text{for} & E<\mu\\
0 & \text{for} & E>\mu
\end{array}\right. \,,
\end{equation}
where $g_d$ is the number of spin degrees of freedom, $\mu$ is the chemical potential (that for lower temperatures approximates the Fermi energy $\mu\xrightarrow{T\rightarrow0}E_F$), and Planck and Boltzmann constants, $\hbar$ and $k_B$, were set equal to unity.
Reporting about Eq.~\eqref{T_General}, the pressure and the energy density for a DFG are given by
\begin{equation}
\label{T_dfg}
\begin{split}
 \mathcal{\bar P}_d=& g_d \,\frac{m^4}{16\pi^2}\pc{\frac{\sqrt{1+\chi^2}}{\chi^4}\pr{\frac{2}{3}-\chi^2}+\ln{\frac{1+\sqrt{1+\chi^2}}{x}}}\,,\\
 \bar\rho_d=& g_d \, \frac{m^4}{16\pi^2}\pc{\frac{\sqrt{1+\chi^2}}{\chi^4}(2+\chi^2)-\ln{\frac{1+\sqrt{1+\chi^2}}{\chi}}}\,,
\end{split}
\end{equation}
where $\chi\equiv (m/q_F) a$, $m$ is the particle mass, and $q_F$ is the comoving Fermi momentum.

Assuming the fluid behaves as a test-fluid under perturbations due to a background scalar potential, $\phi$, the corresponding continuity and Euler equations would be given by
\begin{subequations}
\label{conserv_dfg}
\begin{equation}
 \d{\delta\rho_d}{\eta}=-\frac{m^4}{3\pi^2}\frac{\sqrt{1+\chi^2}}{\chi^4}\pr{\theta_d-3\dot\phi}-\h\;\delta\rho_d\frac{4+3\chi^2}{1+\chi^2}\,,
\end{equation}
\begin{equation}
 \dot\theta_d+\sigma_d\,k^2\,+\h\frac{\theta_d}{1+\chi^2}-\phi \,k^2\,=\delta\rho_d\frac{\,k^2\,\chi^4\pi^2}{m^4(1+\chi^2)^{3/2}}\,.
\end{equation}
\end{subequations}
where we have identified the DFG through the index $d$.
Supposing that the DFG averaged density is equivalent to the averaged density of massive neutrinos, and considering that neutrinos are ultra-relativistic only deep inside the radiation-dominated era, one can simplify the above obtained equations by setting $\chi\ll1$ during the radiation-dominated era (when $\delta\mathcal{P}_d/\delta\rho_d\approx\mathcal{\bar P}_d/\bar\rho_d\approx1/3$) and by $\chi > 1$ after the radiation-to-matter transitory regime (when $\delta\mathcal{P}_d/\delta\rho_d\approx1/3\chi^2$ and $\mathcal{\bar P}_d/\bar\rho_d\approx1/5\chi^2$), which is quantified by $a = a_{eq}$.

\subsection{Radiation-to-matter transitory regime at very large scales}

\begin{figure}
\centering
\includegraphics[width=14cm]{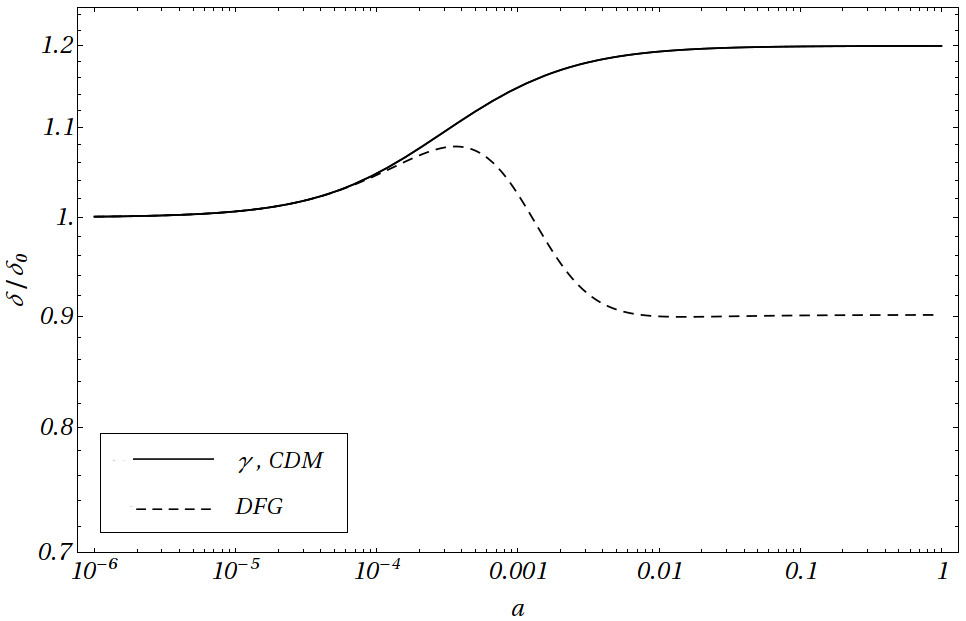}
\caption{Growing mode of the density contrast $\delta$ for $DFG$, radiation and $CDM$ at super-horizon approximation (i. e $k\eta\ll 1$) in dependence on the scale parameter, $a$.
The plots correspond to analytical solutions in the longitudinal gauge for radiation- and matter-dominated universes.}\label{super_dfg_dig}
\end{figure}

For $k\eta\ll 1$ and with vanishing $\sigma$, Eq.~\eqref{conserv_dfg} becomes	
\begin{subequations}\label{eq_super}
\begin{equation}
 \d{\ln\pc{{\chi(\bar\rho_d+\mathcal{\bar P}_d)\theta_d}}}{\ln \chi}=-3\,,
\end{equation}
\begin{equation}
 \chi\d{}{\chi}\delta\rho_d-\frac{m^4}{\pi^2}\frac{\sqrt{1+\chi^2}}{\chi^3}\d{\phi}{\chi}+\delta\rho_d\frac{4+3\chi^2}{1+\chi^2}=0\,.
\end{equation}
\end{subequations}
Using the analytical solution for $\phi$ in case of super-horizon scales (c. f. Eq.~\eqref{phi_Super}), one thus obtains
\begin{equation}
 \theta_d=\frac{\pi^2}{m^4}\frac{3}{\sqrt{1+\chi^2}}S_{0}\,,
\end{equation}
\begin{equation}
 \delta_d=8\pc{S_1\frac{\sqrt{1+y}}{y^3}+S_2\frac{16+8y-2y^2}{y^3}+S_{4}}\pc{(2+\chi^2)-\frac{\chi^4}{\sqrt{1+\chi^2}}\mathrm{arccsch}\,\chi}^{-1}\,.
\end{equation}
where $S_{i}$, with $i = 0,\, 1,\, 2,\, 3$, are constants used to independently match preliminarily uncoupled solutions for radiation- and matter-dominated eras.
Although the density contrasts for cold dark matter (CDM) and photons remain the same for the period of radiation-to-matter transition, Fig.~\ref{super_dfg_dig} shows that the density contrast for a DFG does not reach the same growing rate of that obtained for CDM at the end of the matter-dominated era, in case of super-horizon scales.
We shall see that such a behaviour recurrently happens at scales $\gtrsim Mpc /h$, with a non-trivial contributions to the matter power spectrum today.

\subsection{Radiation-dominated era}

For the radiation-dominated era with vanishing $\sigma$, the components of Eq.~\eqref{conserv_dfg} (or Eq.~\eqref{dot_delta} and Eq.~\eqref{dot_theta}) can be approximated by
\begin{equation}
\label{B_r_rd}
 \dot\delta_d = -\frac{4}{3}\theta_d + 4\dot\phi\,,   \qquad\mbox{and}\qquad \dot\theta_d =	 \,k^2\, \frac{1}{4}\delta_d + \,k^2\,\phi\,.
\end{equation}
Using the analytical solution for $\phi_{\gamma}$ from Eq.~\eqref{sol_rad}, one thus obtains
\begin{subequations}\label{sol_dfg_rd}
\begin{equation}
\label{Sol_B_r_rd}
 \delta_d = 2C_1\frac{2(1-\omega^2)\cos\omega+\omega(2-\omega^2)\sin\omega}{\omega^3}+2C_2\frac{2(1-\omega^2)\sin\omega-\omega(2-\omega^2)\cos\omega}{\omega^3}\,,
\end{equation}
\begin{equation}
\label{Sol_B_r_rd2}
 \theta_d = \frac{\sqrt 3 }{2}C_1\frac{(\omega^2-2)\cos\omega-2\omega\sin\omega}{\omega^2} +\frac{\sqrt 3 }{2}C_2\frac{(\omega^2-2)\sin\omega+2\omega\cos\omega}{\omega^2}\,,
\end{equation}
\end{subequations}
where $C_{i}$, with $i = 1,\, 2$, are also fitting constants.
In this case, the DFG perturbation reproduces the same behaviour of radiation perturbations since, for the ultra-relativistic limit, one considers $\chi\ll 1$.
The solutions for the growing mode related to the coefficient $C_2$ correspond to oscillating perturbations. 
It occurs because the non-vanishing pressure has prejudiced the gravitational collapse.

\subsection{Matter-dominated era}

Deep inside the matter-dominated era, when the critical density approximates $\bar\rho_m$,  also with vanishing $\sigma$, the components of Eq.~\eqref{conserv_dfg} become
\begin{equation}
\label{C_dfg_rd}
\dot\delta_d = -\theta_d + 3 \dot\phi\,,\qquad\mbox{and}\qquad \dot\theta_d=\frac{\,k^2\,\delta_d}{3\chi^2}- \,k^2\,\sigma_d+\,k^2\,\psi -\frac{\dot a}{a}\theta_d\,.
\end{equation}
Using the analytical solution for $\phi_m$ from Eq.~\eqref{sol_mat}, one finally obtains
\begin{equation}
\begin{split}
\label{delta_dfg_con_sol} \delta_d=&\quad D_1\frac{12\sqrt{3}c_m^2}{k^9\eta}\pc{\frac{k^4}{5}-12\frac{\,k^2\,}{\eta^2}+96c_m^2}-D_2\frac{\,k^2\,\eta^2}{6}+D_3\cos\frac{\sqrt{3}k}{2c_m\,\eta} - D_4\sin\frac{\sqrt{3}k}{2c_m\,\eta} \\
&-D_2\frac{k^4}{8c_m^2}\pc{\cos\frac{\sqrt{3}k}{2c_m\,\eta}\mathrm{Ci}\frac{\sqrt{3}k}{2c_m\,\eta}+\sin\frac{\sqrt{3}k}{2c_m\,\eta}\mathrm{Si}\frac{\sqrt{3}k}{2c_m\,\eta}}\,,
\end{split}
\end{equation}
and
\begin{equation}
\begin{split}
\label{theta_dfg_con_sol}
 \theta_d=&\left\{D_1\frac{9\sqrt{3}}{k^9}\pc{\frac{6k^4}{\eta^4}-\frac{96\,k^2\,c_m^2}{\eta^2}+\frac{8k^4c_m^2}{15}+ 256c_m^4}-D_3\frac{\sqrt{3}k}{c_m}\sin\frac{\sqrt{3}k}{2c_m\,\eta}-D_4\frac{\sqrt{3}k}{c_m}\cos\frac{\sqrt{3}k}{2c_m\,\eta}\right.\\
&\left.+D_2\frac{\sqrt{3}k^5}{8c_m^3}\pc{\mathrm{Ci}\frac{\sqrt{3}k}{2c_m\,\eta}\sin\frac{\sqrt{3}k}{2c_m\,\eta}
-\mathrm{Si}\frac{\sqrt{3}k}{2c_m\,\eta}\cos\frac{\sqrt{3}k}{2c_m\,\eta}}+D_ 2\frac{\,k^2\,\eta}{c_m^2}\pc{\frac{\,k^2\,}{4}+\frac{2\eta^2\,c_m^2}{3}}\right\}\frac{1}{2\eta^2}\,,
\end{split}
\end{equation}
where $c_m = (G\,\rho_0\,m\,\pi\,\Omega_m)/q_F$, and $\mathrm{Si}(z)$ and $\mathrm{Ci}(z)$ are the integral sine and cosine functions.
The term proportional to $\frac{\,k^2\,\eta^2}{6}$ at Eq.~\eqref{delta_dfg_con_sol} shows that the density contrast, $\delta$, grows at the same rate as the expansion parameter, $a$.
The last term between brackets oscillates as $\sin(\sqrt{3}k/(2c_m\,\eta))$ for $\eta<\sqrt{3}k/(2c_m)$, after which, it assumes a logarithmic growing dependence on the conformal time, $\eta$.
Turning to the solution for the fluid velocity divergence, $\theta$, described by Eq.~\eqref{theta_dfg_con_sol}, the terms inside the first brackets describe damped oscillations which are similar to those parameterized by $\sin(\sqrt{3}k/(2c_m\,\eta))$ at Eq.~\eqref{delta_dfg_con_sol}.
For $\eta<\sqrt{3}k/(2c_m)$, the decreasing amplitude is proportional to the inverse of the expansion parameter, i. e. to $1/a$,  after which the oscillating behaviour is completely suppressed.
The last term in the brackets presents a linear growing dependence on the conformal time, i. e.  $\eta\propto\sqrt a$.

In general lines, the perturbations described by Eqs.~\eqref{delta_dfg_con_sol} and \eqref{theta_dfg_con_sol} reproduce an oscillating behaviour at the beginning of the matter-dominated era up to the point where it can be suppressed by the subsequent growing modes.
The complete solutions for $\delta$ and $\theta$ can be depicted from Figs. \ref{delta_dfg,dm} and \ref{theta_dfg,dm}, from which one can notice that DFG perturbations never reach the same growing rate of CDM perturbations.

\subsection{$\Lambda$-dominated era}

If one assumes a non-relativistic dynamics for the DFG test-fluid at late times, that happens when one sets $\chi \gg 1$ at Eq.~(\ref{T_dfg}), the cosmological constant era can be reproduced by the same dynamics that drives the perturbations into the matter-dominated era, (c. f. Eq.~\eqref{C_dfg_rd} with $a$ given by Eq.~\eqref{a(eta)}).
Using the analytical solution for $\phi_\Lambda$ from Eq.~\eqref{sol_const}, one thus obtains the density contrast
\begin{equation}
\begin{split}
\label{delta_dfg_con_sol_ld}
\delta_d=&3F_1\,c_\chi^2(b-\eta)+F_3\cos\pc{\frac{\pi}{2}\frac{\eta(2b-\eta)}{c_\Lambda^2}}+F_4\sin\pc{\frac{\pi}{2}\frac{\eta(2b-\eta)}{c_\Lambda^2}}\\
&+3c_\Lambda\ph{F_2+F_1\,c_\chi^2}\ph{\cos\pc{\frac{\pi}{2}z^2}\mathrm{Fc}\pc{z}+\sin\pc{\frac{\pi}{2}z^2}\mathrm{Fs}\pc{z}}\\
&+\frac{c_\Lambda^3}{\pi}\ph{F_2\,k^2\,-9F_1}\ph{\cos\pc{\frac{\pi}{2}z^2}\mathrm{Fs}\pc{z}-\sin\pc{\frac{\pi}{2}z^2}\mathrm{Fc}\pc{z}}\,,
\end{split}
\end{equation}
where $z\equiv(b-\eta)/c_\Lambda$, $c_\Lambda^2\equiv\sqrt{3}\,c\,m\,\pi/k\,q_F$, $c_\chi\equiv c\:m/q_F$, $\mathrm{Fs}(z)$ and $\mathrm{Fc}(z)$ are respectively the Fresnel integral sine and cosine functions, and $c$ and $b$ are constants respectively given by
\begin{equation}
\label{byc}
c=\sqrt{\cfrac{3}{8\pi G\rho_0\Omega_\Lambda}}\,,  \qquad\qquad b=3\sqrt{\cfrac{3}{8\pi G\rho_0}}\pr{\cfrac{1}{\Omega_m^{1/3}\Omega_\Lambda^{1/6}}-\cfrac{4\Omega_{\gamma}^{1/2}}{\Omega_m}}\,,
\end{equation}
In correspondence with the above density contrast, the solution for the fluid velocity divergence is given by
\begin{equation}
\begin{split}
\label{theta_dfg_con_sol_ld}
 \theta_d=&-9 F_1 (\eta-b)^2+F_3\frac{\pi}{c_\Lambda}z\cos\pc{\frac{\pi}{2}\frac{\eta(2b-\eta)}{c_\Lambda^2}}-F_4\frac{\pi}{c_\Lambda}z\sin\pc{\frac{\pi}{2}\frac{\eta(2b-\eta)}{c_\Lambda^2}}\\
 &-\ph{\cos\pc{\frac{\pi}{2}z^2}\mathrm{Fc}\pc{z}+\sin\pc{\frac{\pi}{2}z^2}\mathrm{Fs}\pc{z}}\ph{F_2\,k^2\,-9F_1}c_\Lambda^2z\\
 &+\ph{\cos\pc{\frac{\pi}{2}z^2}\mathrm{Fs}\pc{z}-\sin\pc{\frac{\pi}{2}z^2}\mathrm{Fc}\pc{z}}\ph{F_2\frac{3\pi}{c_\Lambda}+F_1\sqrt{3}kc_\Lambda c_\chi}c_\Lambda z\,.
\end{split}
\end{equation}
These solutions correspond to decreasing functions that result on decaying modes of perturbation, an effect that match the dynamics ruled by the negative pressure of the dominant cosmological background.

\section{Boltzmann equation solutions for a DFG}

In certain sense, the thermodynamics of a DFG is driven by the cosmological behaviour of the chemical potential, $\mu\sim \mu(a)$.
However, depending on its effective contribution to the fluid dynamics, which sometimes is not relevant, additional contributions to the analytical characteristic of the distribution function are usually discarded by setting $\mu \approx 0$.
That is not the case of our approach, since we are interested in depicting the cosmological behaviour of a DFG.
The simplest way to consistently include the chemical potential in the following calculations is through the zeroth-order Boltzmann equation for a non-interacting fluid (c. f. Eq.~\eqref{PSI_dot_general}) as
\begin{equation}
0=\D{\bar f}{\eta}=\dot a\d{\bar f}{a}\propto\d{}{a}\pr{\frac{E-\mu}{\bar{T}}}\,,
\end{equation}
from which the last equality sets the constraint among the energy, $E$, the temperature, $T$, and the chemical potential, $\mu$.

The condition given by the above equation results in the following constraint for the dependence of the distribution function on the temperature,
\begin{equation}
\left.\d{f}{T}\right|_{\bar T}=\frac{aE}{p}\,\frac{\mu-E}{\bar T}\d{\bar f}{q}=\frac{\epsilon}{q}\,\frac{\nu-\epsilon}{\bar T}\d{\bar f}{q}\,,
\end{equation}
with $\nu = a\mu$, a kind of comoving parametrization of the chemical potential.
It allows one to identify the first-order perturbation coefficient, $\Psi$, that appears into Eq.~\eqref{Psi_dfs_pre}, as
\begin{equation}\label{psi_dfg}
\Psi(x^{i}, n_{j}, \eta) = \frac{\epsilon(\nu-\epsilon)}{q^2}\d{\ln \bar f}{\ln q} \, \Delta(x^{i}, n_{j}, \eta)\,,   \qquad\qquad \end{equation}
where $\Delta\equiv \delta T/\bar T$.
By substituting the above expression for $\Psi$ into Eq.~\eqref{PSI_dot_general}, the Boltzmann equation can be reduced to
\begin{equation}\label{bolt_dfg_n}
\d{}{\eta}\pr{\frac{\epsilon(\nu-\epsilon)}{q}\d{\bar f}{q}\,\Delta} + i (\vec{k} \cdot \hat{n}) (\nu-\epsilon)\d{\bar f}{q}\,\Delta + \pr{q\dot\phi - i (\vec{k} \cdot \hat{n}) \epsilon\,\phi}\d{\bar f}{q} = 0,
\end{equation}
where $\Delta(x^{i}, n_{j}, \eta)$ is read from ${\mathcal D}(k^{i}, n_{j}, \eta)$ in the Fourier $k$-space as
\begin{equation}
\Delta(x^{i},n_{j},\eta) = \frac{1}{(2\pi)^{3}}
\int{dk^{3}\, \exp{[i k_{i} x^{i}]}\, {\mathcal D}(k^{i}, n_{j}, \eta)}.
\end{equation}
The dimensionality of the problem can be reduced by noticing that the evolution of Eq.\eqref{bolt_dfg_n} depends on the direction $\hat{n}$ parameterized by the angle related to $\hat{k} \cdot \hat{n} = \cos\varphi$, which is a natural consequence of the isotropy of the homogeneous background.

Performing the $P_{l}$-Legendre expansion with respect to $\cos(\varphi)$, one obtains
\begin{equation}
{\mathcal D} (k^{i}, n_{j}, \eta) = \sum^\infty_{l=0}(-i)^l(2l+1)\,\Delta_l(k,\eta)\,P_l(\cos(\varphi))\,,
\end{equation}
that results in
\begin{equation}
\Delta_{0} = \frac{1}{4\pi} \int{d\Omega\,{\mathcal D}}, ~~ \Delta_{1} = \frac{i}{4\pi} \int{d\Omega\,\cos\varphi\,{\mathcal D}}~~ \mbox{and}~~\Delta_2=-\frac{3}{8\pi} \int{d\Omega\,\left(\cos^{2}\varphi - \frac{1}{3}\right)\,{\mathcal D}}\,,
\end{equation}
for the first three multipole cofficients.
The corresponding evolution equations for the multipole coefficients, $\Delta_l(k,\eta)$, are obtained through the integration of the Boltzmann equation \eqref{bolt_dfg_n} multiplied by $P_l(\hat k\cdot\hat n) \equiv P_l(\cos{\varphi})$ over the solid angle $d\Omega \equiv d\theta\,d(\cos{\varphi})$, that results in
\begin{equation}\label{expansion_dfg}
\begin{split}
\d{}{\eta}\,\pr{\frac{\epsilon\,(\nu-\epsilon)}{q}\d{\bar f}{q}\,\Delta_0}& + (\nu-\epsilon)\d{\bar f}{q}\,k\,\Delta_1 + q\d{\bar f}{q}\dot\phi =0  \,,\\
\d{}{\eta}\,\pr{\frac{\epsilon\,(\nu-\epsilon)}{q}\d{\bar f}{q}\,\Delta_1} - (\nu-\epsilon)&\d{\bar f}{q}\,\frac{k}{3}\,\pr{\Delta_0-2\Delta_2} + \epsilon\d{\bar f}{q}\,\frac{k}{3}\phi=0 \,,\\
\d{}{\eta}\,\pr{\frac{\epsilon\,(\nu-\epsilon)}{q}\d{\bar f}{q}\Delta_l} + (\nu-\epsilon)\,\d{\bar f}{q}\,\frac{k}{2l+1}&\pc{(l+1)\,\Delta_{l+1}-l\,\Delta_{l-1}}=0 \,,\qquad\text{for}\qquad l\geq2\,.
\end{split}
\end{equation}

The perturbations for the stress-energy tensor from Eq.~\eqref{T_General} can thus be rewritten as
\begin{equation}
\begin{split}\label{variables}
\delta \rho_d=  \frac{\Delta_0}{2\pi^2a^{4}}&\int dq\:q\epsilon^2\,(\nu-\epsilon)\, \d{\bar f}{q}\,, \\
\delta\mathcal{P}_d = \frac{\Delta_0}{6\pi^2a^{4}}&\int dq\:q^3(\nu-\epsilon) \d{\bar f}{q}\,,\\
\theta_d(\bar\rho_d + \mathcal{\bar P}_d)=&\frac{\Delta_1}{2\pi^2a^{4}}\int dq\:q^2\,\epsilon\,(\nu-\epsilon) \d{\bar f}{q}\,,\\
\sigma_d(\bar\rho_d + \mathcal{\bar P}_d)=&-\frac{\Delta_2}{3\pi^2a^{4}}\,\int dq\:q^3\,(\nu-\epsilon)\, \d{\bar f}{q}\,,
\end{split}
\end{equation}
and finally, after integrating Eqs.~\eqref{expansion_dfg} over $q$ and using the results from Eq.~\eqref{variables}, one can resolve the system of coupled equations and find the time dependence of the DFG perturbations.
From such results we shall verify that it is possible to analytically depict the DFG transitory regimes, from relativistic to non-relativistic dynamics.

\subsection{Ultra-relativistic DFG during the radiation-dominated era}

When particles are ultra-relativistic, i. e. $\epsilon = q$, and $\nu$ is set equal to zero, the above results should be reduced to that obtained for massless particles \cite{Ma94,Pas06}.
Differently from which is observed when one sets the ultra-relativistic approximation to $f$, the spectrum does not remain Planckian when $\nu \neq 0$.
The $q$-integration can be performed separately and the dependence on $x^{i}$, $n_j$ and $\eta$ is kept active.
The cosmological behaviour of the temperature depicted from Eq.~\eqref{B} is given by  $T(\eta) \sim T_{0}\,m/[a(\eta)\,q_F]$ since the energy, $E$, is approximated by its ultra-relativistic dependence on the momentum ($\sim p$) through the relation
\begin{equation}\label{approx1}
\frac{E-\mu}{T}\approx q_F \frac{q - q_F}{m \, T_0},
\end{equation}
where $T_0$ is the fluid temperature at present.
By performing the quoted integrations for Eqs.~\eqref{expansion_dfg}, the $\Delta_l$ multipole evolution equations become
\begin{equation}\label{mai}
\dot\Delta_0 + k\Delta_1 - \dot\phi \,c_F = 0,
\end{equation}
\begin{equation}\label{mai1}
\dot\Delta_1 - \frac{k}{3}\pr{\Delta_0-2\Delta_2} - \frac{k}{3}\phi \,c_F = 0,
\end{equation}
and
\begin{equation}\label{mai2}
\dot\Delta_l + \frac{k}{2l+1}\pc{(l+1)\Delta_{l+1}-l\Delta_{l-1}} = 0\,, \qquad\qquad\mathrm{for}\quad l\geq 2,
\end{equation}
where $c_F \equiv q_F^2/(m\, T_0 \ln(2))$.
By considering the analytical solution for $\phi_{\gamma}$ from Eq.~\eqref{sol_rad}, and assuming that the contribution from higher order multipoles $\Delta_l$, with $l \geq  2$, are suppressed, the above equations lead to
\begin{equation}
\Delta_0=\frac{C_1\,c_F}{\omega^3}\pc{\cos\omega\pr{1-\omega^2}+\omega\sin\omega}
+\frac{C_2\,c_F}{\omega^3}\pc{\sin\omega\pr{1-\omega^2}-\omega\cos\omega}+C_3\cos\omega-C_4\sin\omega\,,
\end{equation}	
and
\begin{equation}
\Delta_1=-\frac{C_1\,c_F}{\sqrt{3}\omega^2}\pc{\cos\omega+\omega\sin\omega}
-\frac{C_2\,c_F}{\sqrt{3}\omega^2}\pc{\sin\omega-\omega\cos\omega}
+\frac{C_3}{\sqrt{3}}\sin\omega+\frac{C_4}{\sqrt{3}}\cos\omega\,.
\end{equation}

The above results can be substituted into the integrals from Eqs.~\eqref{variables} in order to give
\begin{equation}
\delta_d\approx2^4\pi^3\,\ln(2)\,\frac{m\,T_0}{q_F^2}\,\Delta_0,
\end{equation}
\begin{equation}
\frac{\delta\mathcal{P}_d}{\delta\rho_d}\approx\frac{1}{3},
\end{equation}
\begin{equation}
\theta_d\approx12\pi^3\,\ln(2)\,\frac{m\,T_0}{q_F^2}\,k\,\Delta_1,
\end{equation}
and, eventually,
\begin{equation}
\sigma_d\approx-2^3\pi^3\,\ln(2)\,\frac{m\,T_0}{q_F}\,\Delta_2,
\end{equation}
that reproduce the results from Eqs.~\eqref{Sol_B_r_rd} and \eqref{Sol_B_r_rd2} obtained through the continuity and Euler equations, i. e. Eqs.~\eqref{delta_dfg_con_sol} and \eqref{theta_dfg_con_sol}, once one has assumed that the DFG behaves as an ideal fluid.

\subsection{Non relativistic DFG during the matter-dominated era}

Deep inside the matter-dominated era, when $\rho_{cr}\approx\bar\rho_m$, the cosmological behaviour of the temperature depicted from Eq.~\eqref{B} is given by  $T(\eta) \sim T_{0}\, a(\eta)^{-2}$ since the energy, $E$, is approximated by its non-relativistic dependence on the momentum ($\sim p^2/(2m)$) through the relation
\begin{equation}\label{approx2}
\frac{E - \mu}{T}\approx\frac{q^2 - q_F^2}{2m \,T_0}.
\end{equation}
By evaluating the quoted integrations of Eqs.~\eqref{expansion_dfg} over the momentum $q$, and assuming that $\Delta_l \sim 0$, for $l \geq  2$, the multipole evolution equations can be written as
\begin{equation}
\label{GDF_mat}
\dot\Delta_0+\frac{k}{\chi}\,\Delta_1 - c_F \dot\phi=0,
\end{equation}
\begin{equation}
\label{GDF_mat1}
\dot\Delta_1-\frac{k}{3\chi}\,\pr{\Delta_0-2\Delta_2} -\frac{k}{3}\, \chi\,c_F\phi=0,
\end{equation}
and
\begin{equation}
\label{GDF_mat2}
\dot\Delta_l + \left(\frac{k}{2l+1}\right) \frac{1}{\chi}\,\pc{(l+1)\Delta_{l+1}-l\Delta_{l-1}}=0\,,\qquad\qquad\mathrm{for} \quad l \geq 2.
\end{equation}
By substituting the analytical solution for $\phi_m$ from Eq.~\eqref{sol_mat} into Eqs.~(\ref{GDF_mat}-\ref{GDF_mat1}), one obtains
\begin{equation}
\begin{split}
 \Delta_0=&\:\:\:\:\:\:D_1\,c_F\frac{4\sqrt{3}c_m^2}{k^9\eta}\pc{\frac{k^4}{5}-12\frac{\,k^2\,}{\eta^2}+96c_m^2} +D_3\cos\frac{\sqrt{3}k}{2c_m\,\eta} -D_4\sin\frac{\sqrt{3}k}{2c_m\,\eta} \\
         &-D_2\,c_F\frac{k^4}{24c_m^2}\pc{\cos\frac{\sqrt{3}k}{2c_m\,\eta}\mathrm{Ci}\frac{\sqrt{3}k}{2c_m\,\eta}
         +\sin\frac{\sqrt{3}k}{2c_m\,\eta}\mathrm{Si}\frac{\sqrt{3}k}{2c_m\,\eta}} - D_2\,c_F\frac{\,k^2\,\eta^2}{18}\,,
\end{split}
\end{equation}
\begin{equation}
\begin{split}
 \Delta_1=&D_1\,c_F\frac{\sqrt{3}c_m}{k^{10}}\pc{\frac{6k^4}{\eta^4}-\frac{96\,k^2\,c_m^2}{\eta^2}+\frac{8k^4c_m^2}{15}+256c_m^4}  -\frac{D_3}{\sqrt{3}}\sin\frac{\sqrt{3}k}{2c_m\,\eta}-\frac{D_4}{\sqrt{3}}\cos\frac{\sqrt{3}k}{2c_m\,\eta}\\
 +&D_2\,c_F\frac{k^4/\sqrt{3}}{24c_m^2}\pc{\mathrm{Ci}\frac{\sqrt{3}k}{2c_m\,\eta}\sin\frac{\sqrt{3}k}{2c_m\,\eta}-\mathrm{Si}\frac{\sqrt{3}k}{2c_m\,\eta}\cos\frac{\sqrt{3}k}{2c_m\,\eta}}+D_ 2c_F\frac{k\eta}{9c_m}\pc{\frac{\,k^2\,}{4}+\frac{2\eta^2\,c_m^2}{3}}\,,
\end{split}
\end{equation}
where $c_m \equiv (G\,\rho_0\,m\,\pi\,\Omega_m)/q_F$, as it was previously defined.
Evaluating the integrations prescribed by Eqs.\eqref{variables} and using the non-relativistic approximation for the DFG through  Eq.~\eqref{approx2}, the perturbations for the stress-energy tensor are given by
\begin{equation}
\label{pert_GDF_no}
\delta_d\approx 12 \pi^3\ln(2)\frac{m\,T_0}{q_F^2}\Delta_0,
\end{equation}
\begin{equation}
\frac{\delta\mathcal{P}_d}{\delta\rho_d}\approx\frac{1}{3\chi^2},
\end{equation}
\begin{equation}
\theta_d\approx 12 \pi^3\ln(2)\frac{m\,T_0}{q_F^2}\frac{\Delta_1}{\chi},
\end{equation}
and, when required,
\begin{equation}
\label{pert_GDF_no2}
\sigma_d\approx8\pi^3\ln(2)\frac{m\,T_0}{q_F^2}\frac{\Delta_2}{\chi}.
\end{equation}
Again, the above solutions reproduce the same behaviour of the solutions from Eqs.~\eqref{delta_dfg_con_sol} and \eqref{theta_dfg_con_sol} for an ideal fluid  approach.
One should notice that the non-null perturbation on the averaged pressure, as obtained through the ideal fluid approach, explains some small deviations from pure cold matter perturbations, for which the pressure should vanish.

\subsection{Non-relativistic DFG during the $\Lambda$-dominated era}

As we have stated in the previous section, when one assumes the non-relativistic dynamics for a DFG test-fluid at late times by setting $\chi \gg 1$ at Eq.~(\ref{T_dfg}), the dynamics for the cosmological constant era can be reproduced by the same equations of matter perturbations.
In this case, it is reproduced by Eqs.~(\ref{GDF_mat}-\ref{GDF_mat2}).
By following the same procedure as for radiation- and matter-dominated eras, i. e. using the analytical solution for $\phi_\Lambda$ from Eq.~\eqref{sol_const} into Eqs.~(\ref{GDF_mat}-\ref{GDF_mat2}), with $\Delta_l \sim 0$, for $l \geq  2$, one obtains
\begin{equation}
\begin{split}
\Delta_0=&F_1\,c_F\,c_\chi^2(\eta-b)+F_3\cos\pc{\frac{\pi}{2}\frac{\eta(2b-\eta)}{c_\Lambda^2}}+F_4\sin\pc{\frac{\pi}{2}\frac{\eta(2b-\eta)}{c_\Lambda^2}}\\
          &+\ph{\cos\pc{\frac{\pi}{2}z^2}\mathrm{c}\pc{z}+\sin\pc{\frac{\pi}{2}z^2}\mathrm{s}\pc{z}}c_Fc_\Lambda\ph{F_2+F_1\,c_\chi^2}\\
          &+\ph{\cos\pc{\frac{\pi}{2}z^2}\mathrm{s}\pc{z}-\sin\pc{\frac{\pi}{2}z^2}\mathrm{c}\pc{z}}c_F\ph{F_2\,c_\chi^{3/2}\sqrt{\frac{k\pi}{\sqrt{3}}}-F_1\,c_\Lambda^3\frac{3}{\pi}}\,,
\end{split}
\end{equation}
\begin{equation}
\begin{split}
 \Delta_1=&-F_1\,c_F\,c_\chi^2\frac{\sqrt{3}}{\pi}(b-\eta)+\frac{F_3}{\sqrt{3}}\cos\pc{\frac{\pi}{2}\frac{\eta(2b-\eta)}{c_\Lambda^2}}-\frac{F_4}{\sqrt{3}}\sin\pc{\frac{\pi}{2}\frac{\eta(2b-\eta)}{c_\Lambda^2}}\\
          &-\ph{\cos\pc{\frac{\pi}{2}z^2}\mathrm{Fc}\pc{z}+\sin\pc{\frac{\pi}{2}z^2}\mathrm{Fs}\pc{z}}\ph{F_2\,c_\chi^{3/2}\frac{\sqrt{k\pi}}{3^{3/4}}-F_1\,c_\Lambda^3\frac{\sqrt{3}}{\pi}}c_F\\
          &+\ph{\cos\pc{\frac{\pi}{2}z^2}\mathrm{Fs}\pc{z}-\sin\pc{\frac{\pi}{2}z^2}\mathrm{Fc}\pc{z}}\ph{F_2+F_1\,c_\chi^2}\frac{c_Fc_\Lambda}{\sqrt{3}}\,.
\end{split}
\end{equation}
with  $c_\Lambda^2 \equiv \sqrt{3}\,c\,m\,\pi / (k\,q_F)$, $c_\chi\equiv c\:m / q_F$, where $c$ and $b$ are given by Eq.~\eqref{byc}.

Since the Eqs.~(\ref{pert_GDF_no}-\ref{pert_GDF_no2}) are still valid, the DFG perturbations during the $\Lambda$-dominated era (obtained using the Boltzmann equation) are equivalent to the solutions obtained using Eqs.~\eqref{delta_dfg_con_sol_ld} and \eqref{theta_dfg_con_sol_ld}.

\section{Numerical solution}

Herewith we shall consider the numerical procedures for treating perturbations valid for test-fluids in a DFG regime.
For such a purpose we have compiled the public code CAMB \cite{CAMB} based on the synchronous gauge scalar perturbations on the flat Friedmann-Robertson-Walker metric with isentropic initial conditions for the stress-energy perturbations.
As one shall notice, the numerical results that we have obtained are in complete agreement with the preliminary analytical studies that we have proposed in the previous sections.

Considering that the perturbed flat Friedmann-Robertson-Walker metric in the synchronous gauge is described by
\begin{equation}
ds^2=a^2\pc{-d\eta^2+\pr{\delta_{ij}+h_{ij}}dx^idx^j}\,,
\end{equation}
the continuity and Euler equations of the stress-energy tensor for ideal fluids can be written as
\begin{equation}
\d{\,\delta\rho}{\eta}=-\dot{\bar\rho}-\pr{\bar\rho+\mathcal{\bar P}}\pr{\theta+\frac{\dot h}{2}+3\h}-3\h\,\delta\rho\pr{1+\frac{\delta\mathcal{P}}{\delta\rho}}\,,
\end{equation}
and
\begin{equation}
\pr{\dot\theta + \sigma \,k^2\, + 4\h\,\theta}\pr{\bar\rho+\mathcal{\bar P}} + \theta \pr{\dot{\bar\rho}+\dot{\mathcal{\bar P}}}=\delta\rho\,k^2\,\frac{\delta\mathcal{P}}{\delta\rho}\,,
\end{equation}
with $h\equiv h_{ii}$.
In analogy to Eq.~\eqref{T_dfg}, one thus obtains
\begin{subequations}\label{syn}
\begin{equation}
\d{\,\delta\rho}{\eta} = -\frac{m^4}{\pi^2}\frac{\sqrt{1+\chi^2}}{3\chi^4}\pr{\theta+\frac{\dot h}{2}} - \delta\rho\,\left(\h\right)\frac{4+3\chi^2}{1+\chi^2}\,,
\end{equation}
and
\begin{equation}
\dot\theta + \sigma \,k^2\, + \theta\left(\h\right)\,\frac{2\chi^2}{1+\chi^2}=\delta\rho\, \frac{\chi^4\,k^2\,\pi^2}{m^4\pr{1+\chi^2}^{3/2}}\,,
\end{equation}
\end{subequations}
which are used in the numerical calculations
We have performed the numerical calculations to obtain the time dependence of the DFG perturbation by introducing the constraints given by Eq.\eqref{syn} into the CAMB code, with $\sigma_d\approx0$.

\subsection{Background and perturbative initial conditions}

To compare DFG with massive neutrino perturbations in the background of the $\Lambda CDM$ model, we have set coincident values for the averaged densities of both fluids as preliminary conditions, at early times (radiation-dominated) and at very late times (matter-dominated era).
To obtain the comoving Fermi momentum $q_F$ for a DFG, we have considered that the ultra-relativistic limit of Eq.~\eqref{T_dfg},
\begin{equation}
\bar\rho_d(\chi\ll 1)\approx\frac{m^4}{8\pi^2}\frac{g_d}{\chi^4}=\frac{g_d}{4\pi^2}\frac{q_F^4}{a^4}\,,
\end{equation}
does not depend on the mass, $m$.
By equating the averaged densities of the equivalent matter components: DFG and neutrinos, i. e. by setting $\bar\rho_d =\bar\rho_\nu$ (assuming three degenerated families for each one of them), one easily verifies that
\begin{equation}
\bar\rho_d(\chi\ll 1)=\bar\rho_\nu(a\ll a_{eq})=3\frac{7}{8}\pr{\frac{4}{11}}^{4/3}\bar\rho_\gamma(a)\,,
\end{equation}
which results in $q_F\approx3.65\cdot10^{-4}\,eV/c$.

By introducing the obtained value into the non-relativistic approximations for the DFG averaged density,
\begin{equation}
\bar\rho_d(\chi\gg1)\approx\frac{m^4}{6\pi^2}\frac{g_d}{\chi^3}=\frac{g_d}{6\pi^2}\frac{q_F^3}{a^3}m\,,
\end{equation}
and following the same equality, $\bar\rho_d =\bar\rho_\nu$, in the limit where $a$ tends to unity at present,
\begin{equation}
\bar\rho_d(\chi\gg1)=\bar\rho_\nu(a\to1)=\rho_{0}\frac{\Omega_\nu}{a^3}\,,
\end{equation}
one can easily deduce the value for an effective neutrino mass given by $m\approx0.24\,eV/c^2$.

Finally, by integrating the isentropic initial conditions over $q$ for the multipole expansion of the massive neutrino perturbations \cite{Ma94}, and using the DFG phase-space distribution, we reproduce the same initial conditions for DFG perturbations as those for massless neutrinos.

\begin{figure}
\centering
\includegraphics[width=8.2cm]{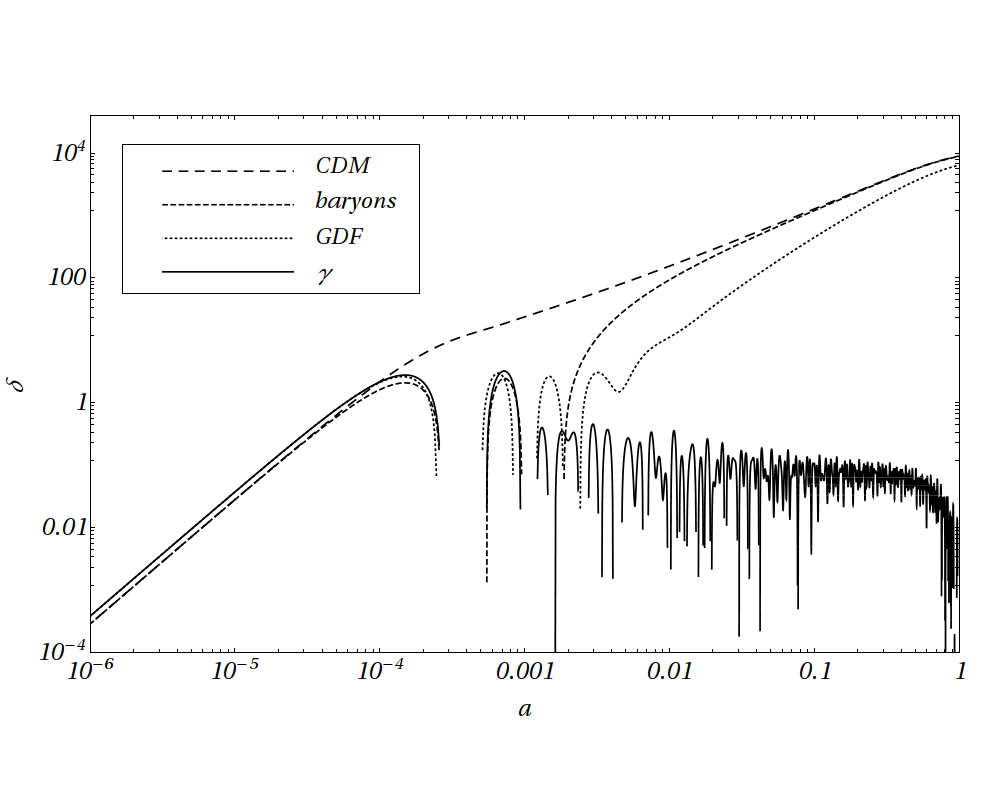}\includegraphics[width=8.2cm]{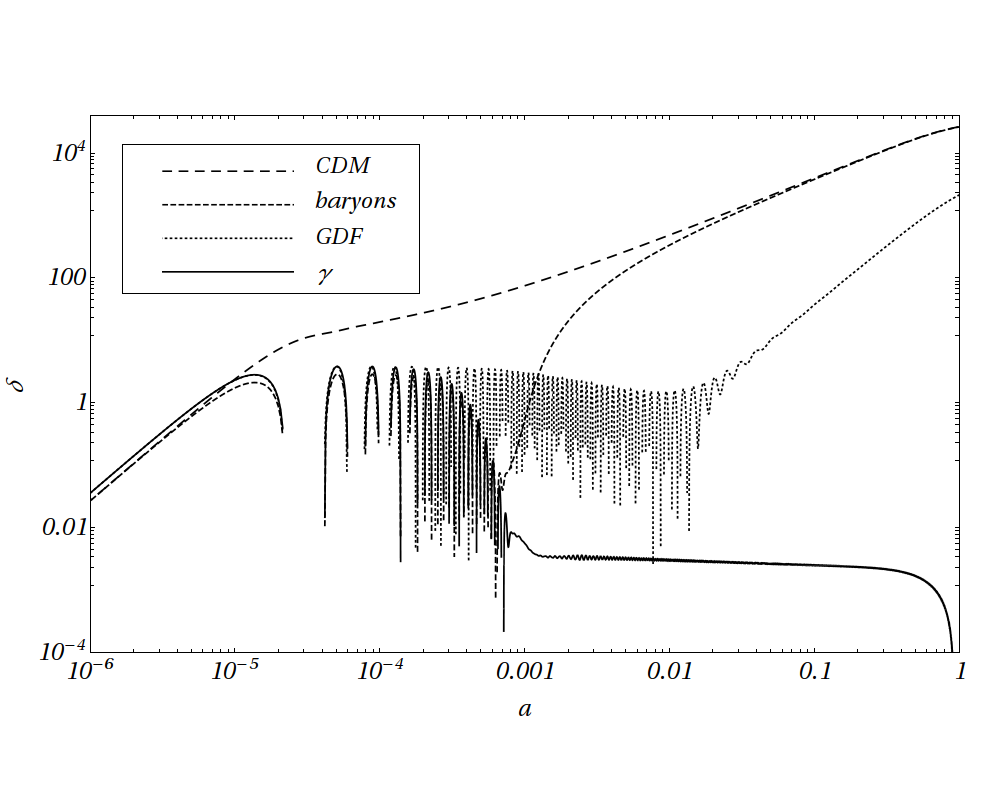}
\caption{Numerical results for the density contrast, $\delta$, (in the synchronous gauge) for $CDM$, baryons, $DFG$ and photons as function of the scale parameter, $a$.
The calculations were performed in the longitudinal gauge for two different scales parameterized by the wavenumber $k$; $k=0.1$ MPc$^{-1}$ (left) and $k=1$ MPc$^{-1}$ (right).}\label{delta_Fig}
\end{figure}

Fig.~\ref{delta_Fig} shows the density contrast, $\delta$, that we have obtained for a DFG.
During the radiation-dominated era $\pr{a\ll a_e\approx2,7\cdot10^{-4}}$ the DFG density contrast, $\delta_d$, has the same oscillating behaviour of radiation perturbations, $\delta_\gamma$, since both fluids have approximately the same evolution equations.
It occurs for a DFG approximated by ultra-relativistic conditions.
During the matter-dominated era the DFG density contrast oscillates up to the beginning of the non-relativistic regime.
After being transposed to the non-relativistic scenario, the DFG density contrast oscillating modes are suppressed and the growing modes reproduce the same behaviour of the CDM perturbation.
Finally, when the Universe passes to the $\Lambda$-dominated era, the DFG density contrast growing modes are suppressed by the negative pressure of the dominant fluid.

Fig.~\ref{deltas_Fig1} allows us to depict the difference between the density contrast of a DFG, for which we have suppressed the higher order multipoles, and the density contrast of massive neutrinos evaluated numerically.
The time dependence of the perturbations for both fluids has two fundamental differences.
Firstly, as $\sigma$ vanishes only for our DFG model, the damped oscillations for the massive neutrino density contrast are not observed for the DFG during the radiation-dominated era (ultra-relativistic period).
In addition, since we have followed the above approach for matching DFG and neutrino energy densities, the neutrino mass for a completely non-relativistic fluid is supposed to be given by $m_\nu \approx 0.46\,eV$, while the DFG was already stated as $m_d \approx 0.24\,eV$.
Due to such a peculiarity, the massive neutrinos become non-relativistic earlier.
These two opposed effects compensated themselves so that the density contrast for massive neutrinos and for a DFG have equivalent amplitudes at present.
The matter power spectrum for $\Lambda CDM + \nu$ and $\Lambda CDM + DFG$ universes are therefore very similar for large and small scales, with a small relative deviation for intermediate scales.

Fig.~\ref{espectro} shows the difference between the matter power spectrum, $P_m(k)$, for the cosmic inventories described at Table \ref{tabla1}.
The matter power spectrum is defined in the Fourier $k$-space as the averaged of the correlation function of the density contrast as
\begin{equation}
\langle\delta^{\dagger}_m(\vec{k},\, \eta) \delta_m(\vec{k}^{\prime},\,
\eta)\rangle = (2 \pi)^3 P_m(k)\, \delta^3(\vec{k} - \vec{k}^{\prime}).
\end{equation}
By observing that $\delta_m = (\sum_{i}\bar{\rho}_{i}\delta_{i})/(\sum_{i}\bar{\rho}_{i})$ (all $i$ matter components) and integrating the above equation over the solid angle
$d\Omega_k$, one can depict an explicit value for $P_m(k)$.
All the resulting matter power spectra were normalized to match one each other at large scales.
Fig.~\ref{espectro} shows a certain level of decreasing on the $\Lambda$CDM power spectrum at small scales.
This is due to the introduction of an additional component into the cosmic inventory, namely massless neutrinos, massive neutrinos or even a DFG test-fluid.
Such a large difference is indeed evident for small scales, i. e. when one compares the density contrast between CDM plus barions and  CDM plus barions modified by additional (and eventually, exotic) components, as it has been discussed in some previous references \cite{Pas06}.

\begin{table}[ht]
\centering
\begin{tabular}{|c|c|c|c|c|}
  \hline
  $\Omega$ & without $\nu$ & massless $\nu$ & massive $\nu$ & DFG\\
  \hline
  \hline
  $\Omega_\Lambda$ & 0.73 & 0.73 & 0.73 & 0.73\\
  $\Omega_\gamma$ & 4.6E-5 & 4.6E-5 & 4.6E-5 & 4.6E-5\\
  $\Omega_b$ & 0.0425 & 0.0425 & 0.0425 & 0.0425\\
  $\Omega_c$ & 0.2+r & 0.2 & 0.2 & 0.2\\
  $\Omega_r$ & 0 & r & 0 & 0\\
  $\Omega_\nu$ & 0 & 0 & r & 0\\
  $\Omega_d$ & 0 & 0 & 0 & r\\
   \hline
\end{tabular}
\caption{Cosmic inventory corresponding to the matter power spectrum of Fig. \ref{espectro}.
It is assumed a flat universe with$\Sigma_i\Omega_i=1$ that results in $r=0.027454$}\label{tabla1}
\end{table}

Just to end up, in what concerns the neutrino degeneracy, one can ascertain that the value of the chemical potential of electron neutrinos ($\mu_e/T_0$) should be close to $-1$ in order to explain with an acceptable confidence level the phenomenology related to the observed variation of deuterium by roughly an order of magnitude \cite{Dolgov02}.
It would induce a variation in the total energy density during the radiation-dominated stage, which is excluded by the smoothness of the CMB.
However, this objection could be avoided if there was a coincidence between different leptonic chemical potentials
such that in different spatial regions they could have the same values but with the interchange of electronic, muonic and/or tauonic chemical potentials \cite{Hu93A}.
For further details and discussions on the energy dependence and effective chemical potential one might address to preliminary calculations \cite{Hu93A,Hu93B} where the main points are that they are not obtrusive to our results.

\section{Conclusions}

The hypothesis of a small fraction of matter components evolving cosmologically as a DFG test-fluid in some dominant cosmological background was evaluated through numerical and analytical approaches.
Our proposal was based on the fact that for any type of fermionic dark matter, which includes relativistic and non-relativistic neutrinos, the average phase-space density cannot exceed the phase-space density of the DFG, which leads to the possibility of forming extended overdense regions of a fluid of degenerate particles.
In the scope of our analysis, we have introduced some analytical procedures for obtaining the evolution of perturbations for some categories of relativistic and non-relativistic test-fluids.
Expressions for the time evolution of density contrasts and fluid velocity divergencies with vanishing anisotropic stress, $\sigma$, were obtained for a gas of massive and degenerate fermionic particles in the background regimes of radiation-, matter- and $\Lambda$-dominated universes, and for super-horizon scales during the radiation-to-matter transition period.
Mathematically, one can depict from the set of Eqs.(\ref{variables}) that the anisotropic stress is proportional to the quadrupole term in the same fashion that the DFG pressure perturbation is proportional to the monopole term.
In this case, the same arguments that one could use for disregarding the contribution from $\sigma$ to the evolution of the perturbations for ultra-relativistic and non-relativistic are analogously maintained.
In case of cosmic neutrinos, one certainly should not discard the contribution due to the anisotropic stress (multipole $l=2$) in case of analyzing CMB and C$\nu$B anisotropies due to scalar perturbations and the C$\nu$B coupling to gravitational waves due to tensorial perturbations \cite{Lat10,Wei04}.
Evidently, that was not the case of the analysis that we have performed.
Anyway, in case of cosmological massive neutrinos, the collision term is suppressed since the collision rate succumbs under the Hubble rate, and neutrinos free-stream.
Specifically concerning the contribution to the matter power spectrum, which comes significatively from non-relativistic neutrinos (during matter-dominated and matter-to-$\Lambda$ present epochs) the $l\geq2$ multipole contribution are completely suppressed.

Turning back to our primary results, during the period when the DFG could be approximated by an ultra-relativistic gas, it was obtained that the density contrast presents an oscillating behaviour similar to that of the photon perturbation during the radiation-dominated era.
Through analogous comparison with matter components, the freezing of DFG components leads to a combined linear plus logarithmic growing mode for the density contrast, similar to which one could obtain for CDM perturbations.
Finally, during the $\Lambda$-dominated era, the DFG perturbation terms are suppressed (as $1/a$) due to the negative pressure of the dominant fluid.

Through numerical calculations, we have quantified the impact of a DFG component into the cosmic inventory on predictions about the matter power spectrum.
It was performed by running the public code CAMB based on the synchronous gauge and comparing the results with the previous analytical studies based on the longitudinal gauge.
For intermediate regimes, one could observe the suppression of the growing adiabatic modes which turns into a smoothly oscillatory behaviour.
Also for the fluid velocity divergence, the transition from ultra-relativistic to non-relativistic regimes was quantified.
In this sense, our results have become a convenient tool for performing preliminary tests for HDM to CDM test-fluid configurations.
Being more specific, the effects quantified through analytical and numerical calculations and their relative impact on the cosmological structure formation were obtained by substituting the massive neutrino fluid by a DFG matter component.
Assuming isentropic initial conditions, the Fermi momentum, $q_F$, and the mass value, $m$, were adjusted in order to give early (radiation-dominated) and late (matter-dominated) times averaged densities that reproduce the cosmological background neutrino densities obtained through standard predictions from the $\Lambda CDM$ model.

\begin{figure}
\centering
\includegraphics[width=8.2cm]{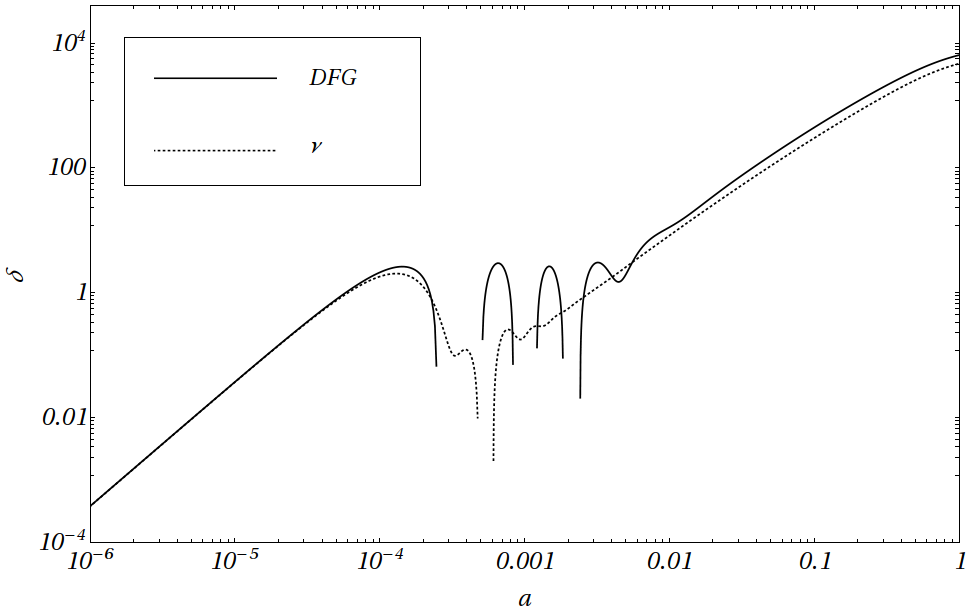}\includegraphics[width=8.2cm]{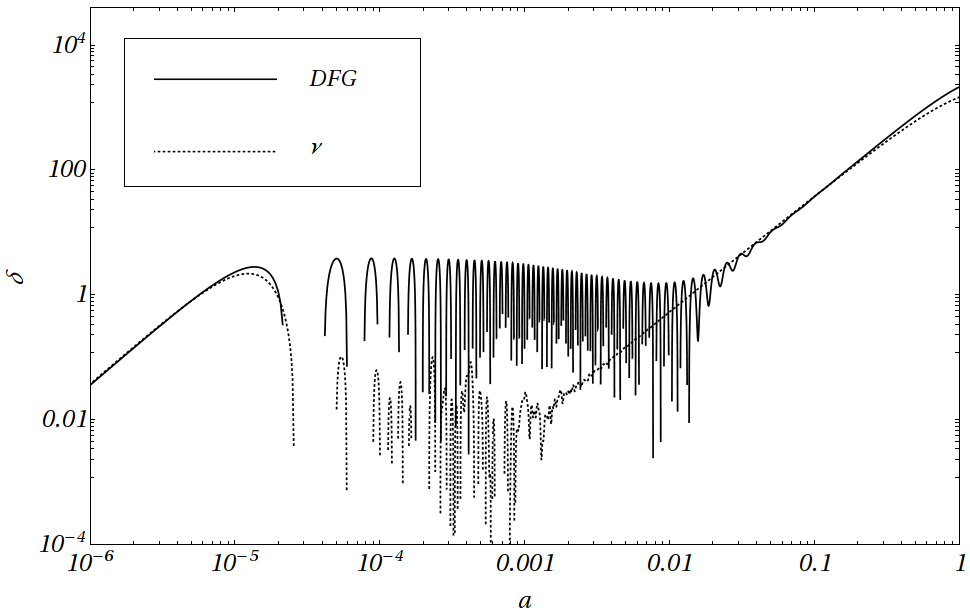}
\caption{Comparison between the massive neutrino density contrast and the $DFG$ density contrast in dependence on the scale parameter, $a$.
It is assumed the same initial and final averaged densities for both fluids (the last two columns on table \ref{tabla1}).
For completeness, the wavenumber here assumed are $k=0.1$ MPc$^{-1}$ (left) and $k=1$ MPc$^{-1}$ (right).}\label{deltas_Fig1}
\end{figure}

\begin{figure}
\centering
\includegraphics[width=14cm]{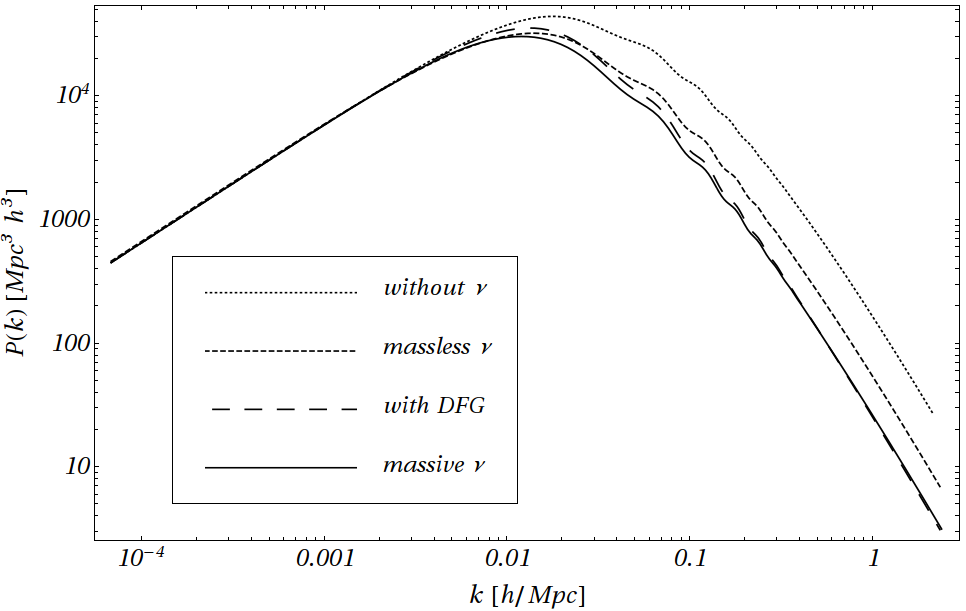}
\caption{Matter power spectrum $P(k)$ according to the cosmic inventory described by the components of table \ref{tabla1} in the $\Lambda CDM$ background scenario.}
\label{espectro}
\end{figure}

Generically speaking, the global evolution of the DFG perturbations obtained by numerical procedures is consistent with the results interpreted from analytical solutions.
During the ultra-relativistic regime, the massive neutrino density contrast presents a damped oscillation while the power is transferred to higher multipoles.
Obviously, through the approximation used for massive neutrinos, they become non-relativistic earlier than a DFG since $m < m_\nu$ for the same averaged densities.
Meanwhile, the density contrast for the DFG and for massive neutrinos are very similar during the non-relativistic regime (c. f. Fig. \ref{deltas_Fig1}), even though it does not reach the growing rate of the CDM density contrast.
The effects on the power spectrum for large and small scales ($k\eta\ll1$ and $k\eta\gg1$ ) are minimal if changing massive neutrinos by DFG.
One of the subtle points of our analysis is that the effects on the matter power spectrum, due to the conversion of massive neutrinos into a DFG fluid in the cosmic inventory, are relevant only for intermediary scales as depicted from Fig.~\ref{espectro}.

To summarize, the reproducibility of our results through the conservation equations derived from the Bianchi identities, through the analytical approximations on the Boltzmann equation, and through the numerical calculations, ratifies the consistency of our analysis.
Once the region of intermediate scales for the matter power spectrum are still open to theoretical speculations, our results could be extended to support the discussion of the formation of galaxy overdense regions of a gas of massive degenerate fermions in hydrostatic and thermal equilibrium at finite temperatures.

\begin{acknowledgments}
The authors would like to thank for the financial support from the Brazilian Agencies CAPES (Master Degree Program - IFGW), FAPESP (grant 08/50671-0) and CNPq (grant 300233/2010-8), and for the hospitality of the Department of Cosmic Rays and Chronology - IFGW - UNICAMP.
\end{acknowledgments}

\pagebreak
\newpage

\pagebreak
\newpage
\begin{figure}
\centering
\includegraphics[width=14 cm]{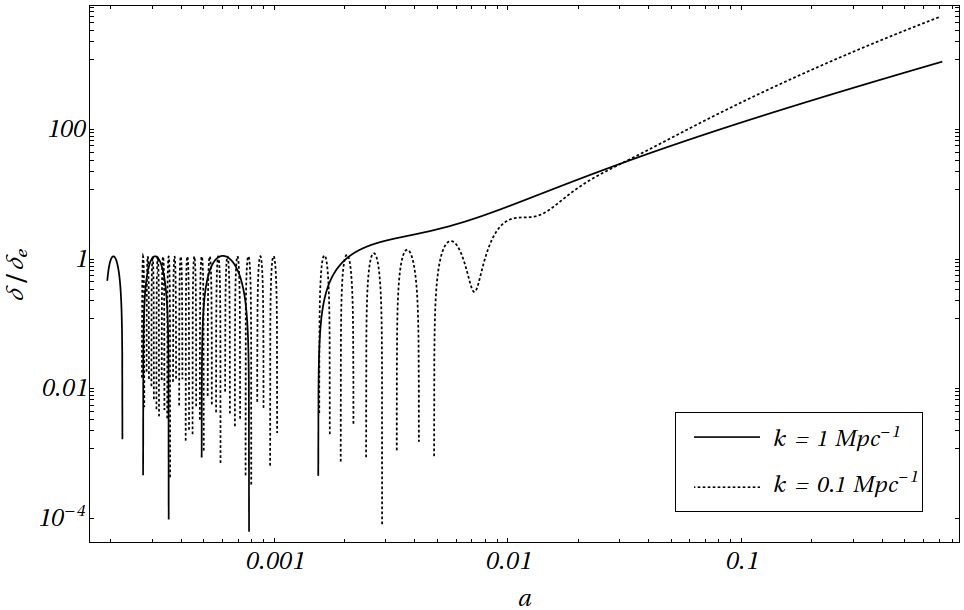}
\caption{Analytical results for the growing mode of the DFG density contrast, $\delta$, during the matter-dominated era as function of the scale parameter, $a$.
The calculations were performed in the longitudinal gauge for two different scales parameterized by the wavenumber $k$; $k=0.1$ MPc$^{-1}$ and $k=1$ MPc$^{-1}$.}\label{delta_dfg,dm}
\end{figure}

\pagebreak
\newpage
\begin{figure}
\centering
\includegraphics[width=14cm]{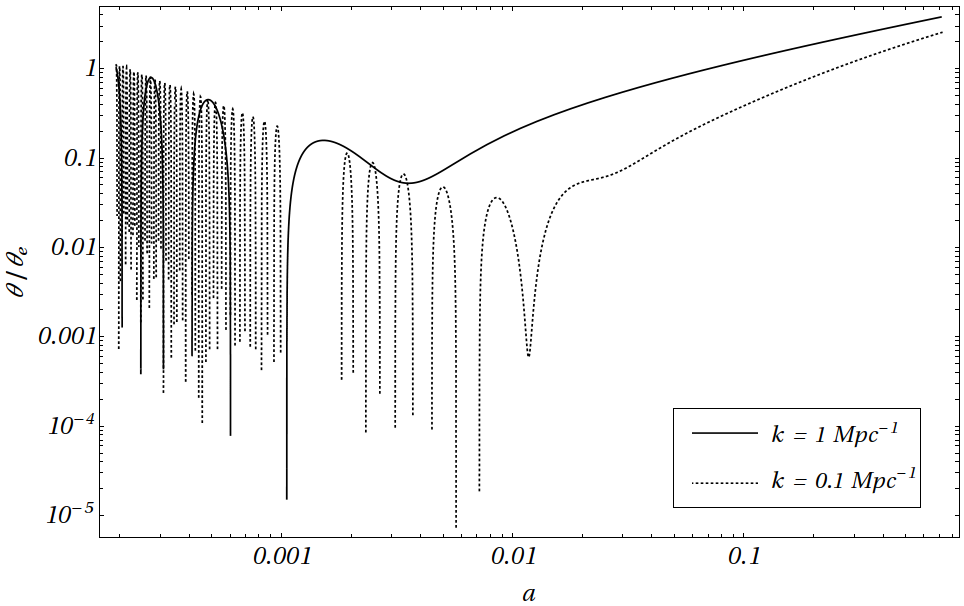}
\caption{Analytical results for the growing mode of the DFG velocity divergence, $\theta$, during the matter-dominated era as function of the scale parameter, $a$.
The calculations were performed in the longitudinal gauge for $k=0.1$ MPc$^{-1}$ and $k=1$ MPc$^{-1}$.}\label{theta_dfg,dm}
\end{figure}

\pagebreak
\newpage

\pagebreak
\newpage

\begin{thebibliography}{99}
\bibitem{Teo2}
Y. B. Zeldovich, A \& A {\bf 5}, 84 (1970).
\bibitem{Teo3}
L. Bergstrom, Rept. Prog. Phys. {\bf 63}, 793 (2000).
\bibitem{Teo4}
G. Bertone, D. Hooper and J. Silk, Phys. Rept. {\bf 405}, 279 (2005).
\bibitem{Teo1}
S. Calchi Novati, Il Nuovo Cimento {\bf B122}, 557 (2007).
\bibitem{Bennet}
C. L. Bennett {\it et al.}, Astrophys. J. Suppl. Ser. {\bf 148}, 1 (2003).
\bibitem{Astropj}
D. Jeong and E. Komatsu, Astrophys. J. {\bf 651}, 619 (2006).
\bibitem{Dodelson}
S. Dodelson, {\em Modern Cosmology: Anisotropies and Inhomogeneities in the Universe}, (Academic Press, New York, 2003).
\bibitem{Dolgov02}
A. D. Dolgov, Phys. Rept. {\bf 370}, 333 (2002).
\bibitem{Boy10}
A. Boyarsky, A. Neronov, O. Ruchayskiy, and I. Tkachev, Phys. Rev. Lett. {\bf 104}, 191301 (2010).
\bibitem{Ma94}
C. P. Ma and E. Bertschinger, Astrophys. J. {\bf 455}, 7 (1995).
\bibitem{Pas06}
J. Lesgourgues and S. Pastor, Phys. Rept. {\bf 429}, 307 (2006).
\bibitem{Ber10}
A. E. Bernardini and O. Bertolami, Phys. Rev. {\bf D81}, 123013 (2010).
\bibitem{Bil97}
N. Bili\'c and R. D. Viollier, Phys. Lett. {\bf B408}, 75 (1997).
\bibitem{Zel81}
A. Dolgov and Ya B. Zel'dovich, Rev. Mod. Phys. {\bf 53}, 1 (1981).
\bibitem{Chandra1939}
S. Chandrasekhar, Ap. J {\bf 74}, 81 (1931); {\em An introduction to the Study of Stellar Structure},  (University of Chicago Press, 1939).
\bibitem{Kremer2002}
C. Cercignani and G. M. Kremer, {\em The relativistic Boltzmann equation: Theory and applications}, in {\em Progress in mathematical physics, Vol. 22}, (Birkhäuser, Basel, Boston, 2002).
\bibitem{Fowler1926}
R. H. Fowler MNRAS {\bf 87}, 114 (1926).
\bibitem{Bernardini2011}
A. E. Bernardini and E. L. D. Perico, JCAP {\bf 01}, 10 (2011).
\bibitem{CAMB}
A. Lewis and A. Challinor, {\em Code for Anisotropies in the Microwave Background}, {\em http://camb.info/};
A. Lewis, Ph.D. thesis, (Cambridge University, 2000), {\em http://cosmologist.info/thesis.ps.gz};
A. Challinor and A. Lasenby, Astrophys. J. {\bf 513}, 1 (1999);
A. Challinor, Phys. Rev. {\bf D62}, 043004 (2000);
A. Lewis and A. Challinor, Phys. Rev. {\bf D66}, 023531 (2002);
A. Challinor and A. Lewis, Phys. Rev. {\bf D71}, 103010 (2005).
\bibitem{Hu93A}
W. Hu and J. Silk, Phys. Rev. Lett. {\bf 70}, 2661 (1993).
\bibitem{Hu93B}
W. Hu and J. Silk, Phys. Rev. {\bf D48}, 485 (1993).
\bibitem{Lat10}
M. Lattanzi, R. Benini and G. Montani, Class. Quant. Grav. {\bf 27}, 194008 (2010).
\bibitem{Wei04}
S. Weinberg, Phys. Rev. {\bf D69}, 023503 (2004).

\end{thebibliography}
\end{document}